\documentclass{optica-article}

\journal{opticajournal} 

\articletype{Research Article}

\usepackage{amsmath}
\usepackage{mathtools}
\usepackage{braket}
\usepackage{lineno}

\begin{document}

\title{Quantum contextuality of complementary photon polarizations explored by adaptive input state control}

\author{Kengo Matsuyama,\authormark{1,*} Ming Ji,\authormark{1,**} Holger F. Hofmann,\authormark{1,***} and Masataka Iinuma\authormark{1,****}}

\address{\authormark{1}Graduate School of Advanced Science and Engineering, Hiroshima University, 
1-3-1 Kagamiyama, Higashi-Hiroshima, 739-8530, Japan}

\email{\authormark{*}matsuyama@huhep.org, \authormark{**}d215181@hiroshima-u.ac.jp, \authormark{***}hofmann@hiroshima-u.ac.jp,  \authormark{****}iinuma@hiroshima-u.ac.jp} 


\begin{abstract*} 
We experimentally investigate non-local contextual relations between complementary photon polarizations by adapting the entanglement and the local polarizations of a two-photon state to satisfy three deterministic conditions demonstrating both quantum contextuality and non-locality. The key component of this adaptive input state control is the variable degree of entanglement of the photon source. Local polarization rotations can optimize two of the three correlations, and the variation of the entanglement optimizes the third correlation. Our results demonstrate that quantum contextuality is based on a non-trivial trade-off between local complementarity and quantum correlations.

\end{abstract*}

\section{Introduction}
Although it seems natural to assume that the values of a physical property are independent of the way in which they are measured, the Bell-Kochen-Specker theorem proves that this assumption cannot be reconciled with the predictions of quantum theory \cite{bell1966problem,kochen1975problem,peres1990incompatible,mermin1990simple,mermin1993hidden,cabello2008experimentally,klyachko2008simple,lapkiewicz2011experimental,pusey2014anomalous,piacentini2016experiment,waegell2017confined,auffeves2019generic}. 
The theoretical predictions of the Bell-Kochen-Specker theorem have been demonstrated in many experiments, including the well-known experimental observation of Bell's inequality violations \cite{PhysRevLett.49.91,ou1988violation,PhysRevLett.81.5039,takesue2004generation,sakai2006spin,ansmann2009violation,kuroda2013symmetric,hensen2015loophole,ballance2015hybrid,giustina2015significant,shalm2015strong,hensen2016loophole,kofler2016requirements,dehollain2016bell,jons2017bright,rosenfeld2017event,big2018challenging,zopf2019entanglement,zhong2019violating,liu2019experimental,proietti2019experimental,paneru2021experimental,qu2021state,ru2022verification}. 
It has been argued that Bell's inequalities are not optimal for the demonstration of non-locality and contextuality because they involve a non-trivial statistical limit. Instead, it may be better to define contextuality in terms of logically necessary conclusions that are then violated with non-vanishing probability or even with certainty in specific quantum systems. The most notable examples of such proofs of non-locality without inequalities are the GHZ paradox \cite{greenberger1990bell,pan2000experimental} and Hardy's paradox \cite{hardy1992quantum,hardy1993nonlocality}, both of which have been confirmed in various experiments \cite{pan2000experimental,PhysRevLett.95.030401,torgerson1995experimental,boschi1997test,boschi1997ladder}.
All of these scenarios indicate that contextuality is a fundamental feature of the non-classical relation between different physical properties. However, the initial quantum state used in these experiments is not defined in terms of its physical properties. Instead, the measurements made to verify the paradox are adapted to the specific quantum coherence of the available input state. Since this initial assumption of quantum coherence makes it difficult to relate the physics of quantum states to the physics of measurement, it may be useful to introduce an operational definition of state preparation in terms of a desired set of measurement outcomes. It should then be possible to define quantum contextuality entirely in terms of deterministic relations between well-defined complementary properties of the physical system.

In the present paper, we explore the possibility of modifying the input state to fit a specific set of statements that define a contextuality paradox. For this purpose, we first identify the physical properties and their desired relations, and then modify the state preparation process until it optimally satisfies the conditions set by these relations. We thus define the initial state entirely in terms of its observable properties, providing a complete operational definition of the physics of state preparation. Our approach has been inspired by the Gedankenexperiment of Frauchiger and Renner \cite{frauchiger2018quantum}, where an entangled state is defined in terms of an interaction intended to provide information about a specific local observable of one of the two systems. However, the quantum correlations obtained as a result of this interaction can also be obtained more directly by an appropriate quantum state preparation process. As we have pointed out in previous work \cite{PhysRevA.107.022208}, the characteristic correlations of the paradox discussed by Frauchiger and Renner uniquely define the initial quantum state in terms of measurement probabilities of zero observed for specific combinations of local outcomes. It is therefore possible to operationally define the state preparation process as a simultaneous suppression of specific outcomes in different measurement contexts. Here, we realize this adaptive state preparation process for polarization entangled photon pairs using a nonlinear crystal from both sides with two-separated pump beams inside of a Sagnac interferometer \cite{PhysRevA.69.013803,kim2006phase,wong2006efficient,lee2016polarization}. This setup allows us to control the degree of entanglement by changing the pump beam polarization, completely preserving the two-photon coherence in the output at all settings. In addition to the degree of entanglement, the local polarization of each photon can be rotated, providing us with three adjustable parameters in the state preparation process. 

As will be shown in the following, it is possible to suppress two of the three outcome probabilities by optimizing only the local polarization rotations. Since the suppression of these probabilities is due to the destructive interference between two output components, the optimal suppression can be found by identifying the local polarization rotation with equal measurement probabilities for the two interfering components. Our results show that this method of outcome suppression is highly efficient, allowing us to optimize the local polarization rotations for a  fixed degree of entanglement. The third outcome probability then depends on the initial degree of entanglement. It is possible to achieve optimal suppression of this probability by varying the degree of entanglement. It should be emphasized that this method of adaptive state control optimizes the state preparation process based on the observed measurement outcomes and requires no further assumptions regarding the quality of the down-conversion source. This approach has three distinct advantages, (a) the optimization of parameters partially compensates the effects of experimental imperfections, improving the performance of noisy sources, (b) quantum state preparation is directly related to the specific correlations relevant for the observation of the contextuality paradox, and (c) the optimization of quantum interference effects can be realized by observing the output probabilities of the components that interfere with each other, providing a practical illustration of complementarity in quantum state preparation.

The goal of adaptive input state control is the optimization of the contextuality paradox associated with the non-zero probability of a specific measurement outcome observed after the three state preparation conditions have been satisfied. We can therefore evaluate the success of the procedure using a contrast function $K$ that relates the residual probabilities of the three outcomes suppressed in adaptive input state control to the non-zero probability that characterizes the paradox. In the present experimental setup, we achieve a contrast of about 0.5 over a wide range of settings controlling the degree of entanglement. The observation of quantum contextuality is thus found to be very robust against changes in the available amount of entanglement. Another interesting aspect of the results is that the maximal contrast $K$ is observed for a degree of entanglement corresponding to an equal balance of local polarization and entanglement. The conditions for the observation of quantum contextuality involve both local and non-local aspects of quantum statistics, possibly revealing new aspects of the relation between local complementarity and entanglement that are hidden by the abstract representation of the initial state as a superposition of specific basis states.

The rest of the paper is organized as follows. Sec. \ref{sec:theory} describes the contextuality paradox and the manner in which it defines adaptive state control. Sec. \ref{sec:experiment} describes the experimental setup and its characteristic properties. In Sec. \ref{sec:results} describes the adaptive state control and the experimental results obtained from it. Sec. \ref{sec:discussion} evaluates the successful observation of quantum contextuality in terms of the contrast $K$. Sec. \ref{sec:conclusion} concludes the paper.

\section{Optimization of quantum contextuality by adaptive input state control}
\label{sec:theory}

Motivated by the Gedankenexperiment introduced by Frauchiger and Renner \cite{frauchiger2018quantum}, we are considering a situation where two separate physical systems satisfy three statements that can be represented by probabilities of zero for three measurement outcomes obtained in different contexts represented by the local measurements of $F_{1/2}$ and $W_{1/2}$, respectively \cite{PhysRevA.107.022208},
\begin{eqnarray}
  P(F_1=0, W_2=a) = 0,
  \label{eqn:deter one} \\
  P(W_1=a, F_2=0) = 0, 
  \label{eqn:deter two} \\
  P(F_1 = 1,F_2 = 1) = 0.
  \label{eqn:deter three}
\end{eqnarray}
Eq. (\ref{eqn:deter one}) and Eq. (\ref{eqn:deter two}) ensure that if the outcome for one system is $W_i=a$, the other system will always show an outcome of $F_j=1$. Since Eq. (\ref{eqn:deter three}) states that a simultaneous detection of $F_j=1$ in both systems is impossible, non-contextual logic suggests that an outcome of $W_1=W_2=a$ for the two systems should be impossible. This impossibility can be expressed by an inequality for the actual probabilities of the corresponding outcomes,
\begin{eqnarray}
  P(a,a) \leq P(0,a) + P(a,0) + P(1,1),
  \label{eqn:criterion of non-contexual logic}
\end{eqnarray}
where $P(x,y)$ represents a joint probability with $x$ as the outcome of $F_1$ or $W_1$ and $y$ as the outcome of $F_2$ or $W_2$. This inequality can be used to confirm the violation of non-contextuality in the presence of experimental imperfections.

As pointed out in \cite{PhysRevA.107.022208}, the three conditions given above correspond to orthogonality relations in the Hilbert space formalism. To satisfy all three conditions, a quantum state $\ket{\psi}$ must be orthogonal to the quantum states representing the measurement outcomes of $\hat{W}_i$ and $\hat{F}_i$,
\begin{eqnarray}
  \braket{0,a | \psi} = 0,
  \label{eqn:condition one} \\
  \braket{a,0 | \psi} = 0,
  \label{eqn:condition two} \\
  \braket{1,1 | \psi} = 0.
  \label{eqn:condition three}
\end{eqnarray}
In the following, the operators $\hat{W}_i$ and $\hat{F}_i$ represent complementary polarization components. This means that the measurement outcomes $0,1$ and $a,b$ are mutually unbiased,
\begin{eqnarray}
  \ket{a} = \frac{1}{\sqrt[]{2}} (\ket{0} - \ket{1})
  \label{eqn:def of a} \\
  \ket{b} = \frac{1}{\sqrt[]{2}} (\ket{0} + \ket{1}).
  \label{eqn:def of b}
\end{eqnarray}
Eqs.(\ref{eqn:condition one})-(\ref{eqn:condition three}) uniquely determine the quantum state $\ket{\psi}$, resulting in a probability of $P(a,a) = 1/12$ \cite{frauchiger2018quantum,PhysRevA.107.022208}. This is a specific version of Hardy's paradox\cite{hardy1992quantum,hardy1993nonlocality}, confirming both quantum contextuality and quantum non-locality. However, the present formulation emphasizes the fact that the initial quantum state is fully defined by experimentally observable conditions given by Eqs.(\ref{eqn:deter one})-(\ref{eqn:deter three}). As shown in \cite{PhysRevA.107.022208}, realizing these conditions will necessarily result in a non-zero probability for the outcome $W_1=W_2=a$. We can therefore develop a method of state preparation that optimizes these conditions in the presence of arbitrary experimental imperfections.

Since the state defined by the conditions in Eqs.(\ref{eqn:condition one})-(\ref{eqn:condition three}) is a partially entangled state, it is necessary to vary the available degree of entanglement. In quantum optics, photon pairs with variable degrees of polarization entanglement can be generated efficiently by parametric down-conversion in a Sagnac interferometer \cite{PhysRevA.69.013803,kim2006phase,wong2006efficient,lee2016polarization}. Since the polarizations of the photons generated in the down-conversion are initially aligned with the optical axes of the non-linear crystal, the initial quantum state is given by
\begin{eqnarray}
  \ket{\psi_0} = \cos{\phi_S} \ket{H,H} - \sin{\phi_S} \ket{V,V},
  \label{eqn:start quantum state}
\end{eqnarray}
where the $\phi_S$ is an experimentally adjustable parameter that controls the degree of entanglement. Any pure state with this degree of entanglement can now be obtained by local polarization rotations. Due to the symmetry of the conditions, we can apply the same polarization rotation $\phi_M$ to both photons, defining the states $\ket{0}$ and $\ket{1}$ as 
\begin{eqnarray}
  \ket{0} = \hat{U}(\phi_M) \ket{H}, \qquad \ket{1} = \hat{U}(\phi_M) \ket{V}.
  \label{eqn:local rotation of polarization}
\end{eqnarray}
Adaptive input state control requires that the rotation angle $\phi_M$ is adjusted to optimally satisfy the conditions Eq. (\ref{eqn:condition one}) and Eq. (\ref{eqn:condition two}) given by $P(0,a)=0$ and $P(a,0)=0$. Experimentally, this procedure can be simplified by making use of Eq.(\ref{eqn:def of a}). The conditions for the quantum state then read
\begin{eqnarray}
  \braket{0,1 | \psi} = \braket{0,0 | \psi},
  \label{eqn:inter01} \\
  \braket{1,0 | \psi} = \braket{0,0 | \psi}.
  \label{eqn:inter10} 
\end{eqnarray}
Experimentally, the coherence is supplied by the initial state $\ket{\psi_0}$, so the optimal rotation angle $\phi_M$ can be found by satisfying the conditions $P(0,0) = P(0,1)$ and $P(0,0) = P(1,0)$. These conditions can always be satisfied exactly, independent of experimental imperfections. Once the local polarization rotations are determined, it is possible to evaluate the actual suppression of the probabilities $P(0,a)$ and $P(a,0)$ by the corresponding measurements. In addition, the value of $P(1,1)$ can be obtained for these local polarization rotations. The optimal suppression of $P(0,a)$ and $P(a,0)$ results in a value of $P(1,1)$ that depends on the parameter $\phi_S$ which is used to control the degree of entanglement. A suppression of $P(1,1)$ thus requires an optimization of $\phi_S$ to adjust the amount of entanglement in the initial state. 

It is interesting to note that only measurements in the $\{0,1\}$-basis are needed for adaptive state control, even though the contextual statistics characterizes the relations between different combinations of $\{0,1\}$ measurements and $\{a,b\}$ measurements. Adaptive state control thus highlights the manner in which the relations between different measurement contexts are defined by quantum coherence in the measurement outcomes \cite{ji2023quantitative}.

\section{Experiment}
\label{sec:experiment}
\subsection{Experimental setup}

\begin{figure}[htbp]
  \centering\includegraphics[width=7cm]{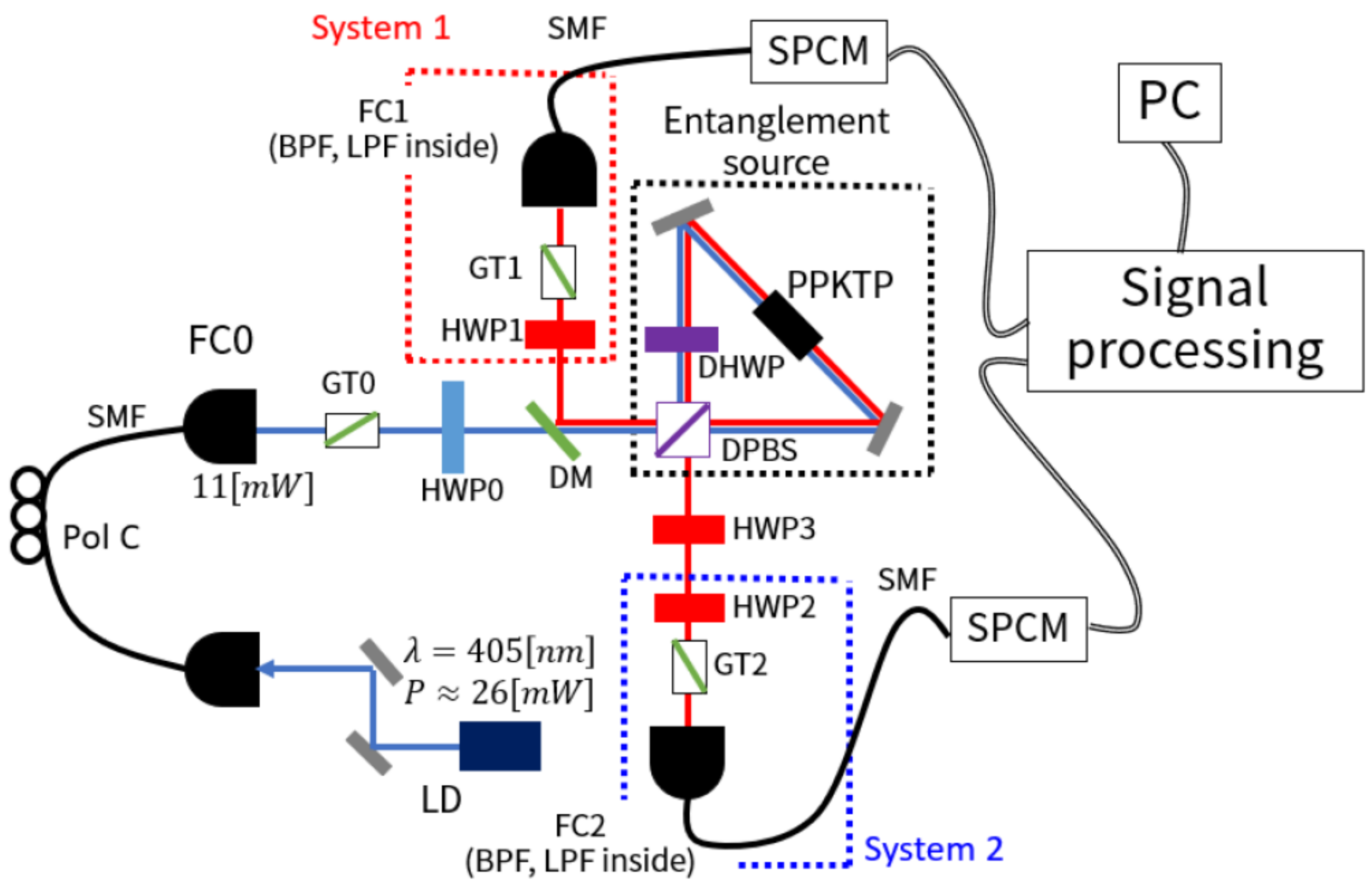}
  \caption{Experimental setup for the adaptive input state control. The pump beam is a laser diode (LD) with a wavelength of 405 nm and its output power is 26 mW. A fiber coupler (FC0) emits the pump beam with power of 11 mW. The pump beam is separated at the double polarization beam splitter (DPBS) into two paths, pumping the periodically poled KTP crystal ($1mm \times 1mm \times 10mm$) from both edges and creating photon pairs. A double half waveplate (DHWP) exchange horizontal and vertical polarizations. The two paths then overlap at the DPBS and the photon pairs are separated there. The two photons are then detected separately by two single photon counting modules(SPCM) after passing through polarization filters, band pass filters (BPF) ($\lambda = 810$ nm, $\delta \lambda = 10$ nm) and long pass filters (LPF) (Transmission:$\lambda > 700 nm$). The detection setup for photon is shown inside the red-dotted box and the blue-dotted box. The output pulses were converted to NIM logic pulses of 30 ns and processed using a NIM logic circuit composed of Logic Level Adapter, Discriminator, and Coincident counting module. The coincidence counts were recorded by a connected PC.}
  \label{fig:setup}
\end{figure}
As shown in Fig. \ref{fig:setup}, the experimental setup consists of a Sagnac-type interferometer used as the source of polarization entanglement, two polarization filters, and two single photon detectors and electronics for signal processing. The pump laser (semiconductor laser, wavelength 405 nm) emits light through a fiber coupler (FC0). The polarization of the pump light can be controlled by the combination of a Glan-Taylor prism (GT0) and a half-wave plate (HWP0) to obtain polarizations from vertical ($V$) polarization to diagonal linear polarization ($P$). The pump beam enters the Sagnac interferometer indicated by a block-dotted box and is separated at a double polarization beam splitter (DPBS). In the clockwise path, the pump beam is converted from $V$ polarization to $H$ polarization by the double half-wave plate (DHWP). Both pump beams interact with the periodically poled KTP (PPKTP) crystal to generate photon pairs with opposite $HV$-polarizations. In the anti-clockwise path, the DPBS separates the photons so that the $V$-polarized photons go leftward and the $H$-polarized photons go downward. In the clockwise path, the $H$-polarized photons go leftward and the $V$-polarized photons go downward. Since the photons from the clockwise path and the anti-clockwise path are indistinguishable, the photon pairs in the output are polarization entangled. The leftward-directed photons are reflected by the dichroic mirror (DM) and measured in system 1 (indicated by the red-dotted box). The $HV$-polarization of the downward-directed photons is flipped from $V$ to $H$ and from $H$ to $V$ by HWP3 to prepare the symmetric quantum state given by Eq. (\ref{eqn:start quantum state}). The  downward-directed photons are then measured in system 2 (surrounded by the blue-dotted box). The photon pairs are measured using combinations of GT1,2 and HWP1,2. Two band pass filters (BPFs) and two long pass filters (LPFs) inside two fiber couplers (FC1 and FC2) eliminate background photons and select photon pairs with $\lambda = 810[nm]$. The photon pairs are detected by two single photon counting modules(SPCMs), PerkinElmer (SPCM-AQR- 14-FC13237-1) and EXCELITAS (SPCM-AQRH-14-FC24360) and their coincidence counts are recorded in a computer. 

In the present setup, the setting of HWP0 determined the control parameter $\phi_S$. If the angle of the HWP0 is set so that the pump beam is $V$-polarized, corresponding to $\phi_S = 0^\circ$, the pump beam passes through the clockwise path only, so that the output photon pairs results in the product state. If the angle of the HWP0 is set so that the pump beam is $P$-polarized, corresponding to $\phi_S = 45^\circ$, the photon pairs are in the maximally entangled state because the pump beam is divided with the almost same intensity into the clockwise and the anti-clockwise paths. Continuously rotating the angle of the HWP0 should therefore provide us with full control over the degree of entanglement generated in the setup. The two HWP2 combine two distinct roles. One is to adjust the local polarization rotation $\phi_M$, which is strictly speaking part of the state preparation, and the other is to switch the measured polarizations between $\{0,1\}$ and $\{a,b\}$. Controlling both settings with only one HWP helps to reduce systematic errors. If the setup is modified for different applications, it may be necessary to perform these two functions using separate HWPs.

To characterize the performance of the entangled photon source, we have evaluated the two-photon visibilities at $\phi_S = 45^\circ$. These visibilities can be given by the correlations
\begin{eqnarray}
  C_i \equiv \left| \frac{N_{++} + N_{--} - N_{+-} - N_{-+}}{N_{++} + N_{--} + N_{+-} + N_{-+}} \right|,
  \label{eqn:def of visibility}
\end{eqnarray}
where $+$ and $-$ are the two possible outcomes for the respective polarization, $N_{++}$ and $N_{--}$ are the count rates for the same polarization, and $N_{+-}$ and $N_{-+}$ are the count rates for opposite polarizations. Evaluating each visibility for the $HV$ and $PM$ polarizations, the experimental results for our entanglement source are
\begin{eqnarray}
  C_{HV} = 0.968 \pm 0.013,
  \label{eqn:visi result on HV} \\
  C_{PM} = 0.935 \pm 0.011.
  \label{eqn:visi result on PM}
\end{eqnarray}
Here, $PM$-polarization refers to diagonal polarizations. The visibility of the $HV$-polarization is limited by imperfection of the optical components used in the setup. In addition, the visibility of $PM$-polarization is limited by the visibility of interference in the Sagnac interferometer. This explains the difference between the visibilities shown in Eq. (\ref{eqn:visi result on HV}) and Eq. (\ref{eqn:visi result on PM}).

\subsection{Entanglement witness and local polarizations}
\label{subsec:the way of evaluate optimal phi_M}
To evaluate the control of entanglement in our setup, it is necessary to characterize the visibilities $C_{HV}$ and $C_{PM}$ for different settings of $\phi_S$. The entanglement generated in the setup can then be evaluated by the entanglement witness given by
\begin{eqnarray}
  W_E \equiv C_{HV} + C_{PM} - 1.
  \label{eqn:def of We}
\end{eqnarray}
Ideally, the reduction of entanglement results in an increase of the local polarization described by the visibility $V_{HV}$,
\begin{eqnarray}
  V_{HV} = \frac{N_H - N_V}{N_H+N_V},
  \label{eqn:def of local visibility}
\end{eqnarray}
where $N_H$ and $N_V$ are the count rates of $H$-polarization and $V$-polarization for a single photon. Since the count rates can be different for the two photons, we obtained the average visibility $V_{HV}$ for both outputs. The relation between the local polarization visibility $V_{HV}$ and the entanglement witness $W_E$ is given by the inequality
\begin{equation}
\sqrt{V_{HV}^2 + W_E^2} \leq 1.
\end{equation}
The limit of 1 is obtained by the ideal state given in Eq. (\ref{eqn:start quantum state}). It is therefore possible to quantify errors in the state preparation process by evaluating the length of the vector $(V_{HV},W_E)$. Similarly, the two photon visibility $C_{HV}$ of the ideal state is always 1. We can thus define two measures that evaluate the performance of our entanglement source. 

\begin{figure}[htbp]
  \centering
  \includegraphics[width=120mm]{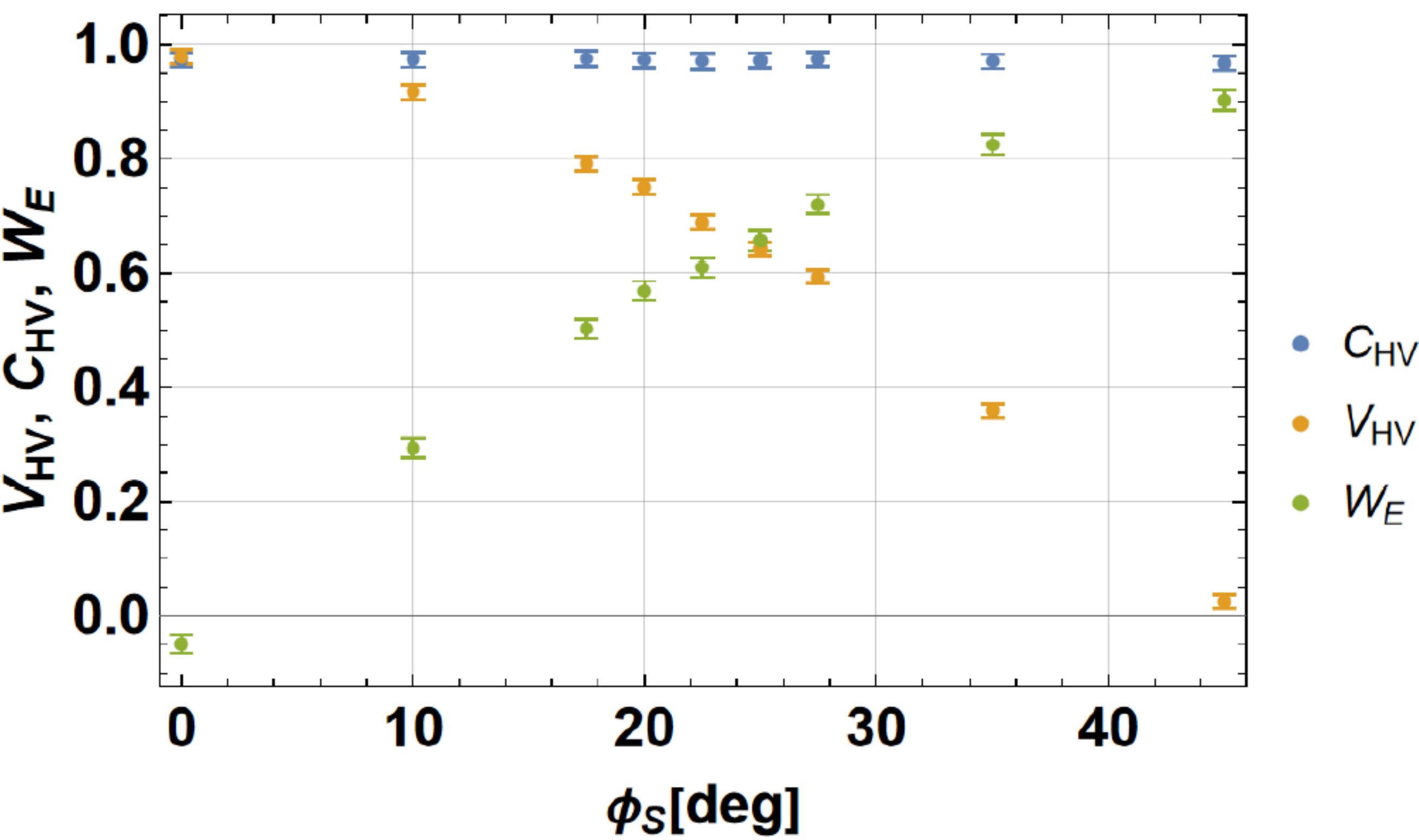}
  \caption{Trade-off relation between the local polarization $V_{HV}$ (yellow dots) and the entanglement witness $W_E$ (green dots) controlled by the parameter $\phi_S$. The blue dots show the correlation $C_{HV}$ of $HV$-polarization. As expected, $\phi_S$ controls the degree of entanglement without changing the correlation $C_{HV}$. See the supplemental material for the raw data obtained in the experiment.}
  \label{fig:experimental result of chv vhv we}
\end{figure}

\begin{table}[htbp]
  \centering
  \caption{Experimental values of $C_{HV}, V_{HV}, W_E$ and $\sqrt{V_{HV}^2 + W_E^2}$ for different settings of the parameter $\phi_S$, including statistical error margins.}
  \label{table:result of the visibilities}
  \footnotesize
  \begin{tabular}{@{}ccccc} \hline
   $\phi_S[deg]$ & $C_{HV}$ & $V_{HV}$ & $W_E$ & $\sqrt{V_{HV}^2 + W_E^2}$ \\ \hline
   $0$ & $0.975\pm 0.013$ & $0.979\pm 0.013$ & $-0.049\pm 0.01$6 & $0.980 \pm 0.013$ \\
   $10$ & $0.974\pm 0.013$ & $0.918\pm 0.013$ & $0.294\pm 0.017$ & $0.963 \pm 0.014$  \\
   $17.5$ & $0.976\pm 0.013$ & $0.793\pm 0.013$ & $0.503\pm 0.017$ & $0.938 \pm 0.014$  \\
   $20$ & $0.973\pm 0.013$ & $0.751\pm 0.012$ & $0.569\pm 0.017$ & $0.943 \pm 0.014$  \\
   $22.5$ & $0.971\pm 0.013$ & $0.689\pm 0.012$ & $0.610\pm 0.017$ & $0.920 \pm 0.015$  \\
   $25$ & $0.973\pm 0.013$ & $0.642\pm 0.012$ & $0.656\pm 0.017$ & $0.918 \pm 0.015$  \\
   $27.5$ & $0.975\pm 0.013$ & $0.594\pm 0.012$ & $0.720\pm 0.017$ & $0.933 \pm 0.015$  \\
   $35$ & $0.971\pm 0.013$ & $0.359\pm 0.011$ & $0.825\pm 0.018$ & $0.900 \pm 0.017$  \\
   $45$ & $0.968\pm 0.013$ & $0.025\pm 0.011$ & $0.903\pm 0.018$ & $0.903 \pm 0.018$  \\ \hline
  \end{tabular} \\
\end{table}

Fig. \ref{fig:experimental result of chv vhv we} shows the experimental results obtained for $V_{HV}$, $C_{HV}$, and $W_E$. The values of $C_{HV}$ are consistent with the result obtained at $\phi_S = 45^\circ$. At $\phi_S = 0^\circ$, the local polarization is maximally defined and $V_{HV}$ is approximately equal to $C_{HV}$. On the other hand, $W_E$ is limited by the visibility of $PM$-correlations, and these depend on quality of interference at the output of the Sagnac interferometer. Therefore $W_E$ stays below the value of $C_{HV}$ at $\phi_S = 45^\circ$, corresponding to the difference between the $HV$ and the $PM$ visibilities in Eqs. (\ref{eqn:visi result on HV}) and (\ref{eqn:visi result on PM}). For intermediate values of $\phi_S$, Fig. \ref{fig:experimental result of chv vhv we} shows the trade-off relation between the local polarization $V_{HV}$ and the entanglement witness $W_E$, where $\phi_S$ controls the degree of entanglement as theoretically predicted. The precise numerical values obtained in the experiment are shown in Table \ref{table:result of the visibilities}. In addition, the values of $\sqrt{V_{HV}^2 + W_E^2}$ are shown to indicate how close the experimental result is to the ideal state. The reason why $\sqrt{V_{HV}^2 + W_E^2}$ shows a significant dependence on $\phi_S$ is that the entanglement witness $W_{E}$ is sensitive to the interference effects at the output of the Sagnac interferometer. As expected, $\sqrt{V_{HV}^2 + W_E^2}$ is equal to $V_{HV}$ at $\phi_S = 0^\circ$ and equal to $W_E$ at $\phi_S = 45^\circ$. The value of $\sqrt{V_{HV}^2 + W_E^2}$ thus drops as the relative contribution of $W_E$ increases.

\section{Optimization of contextual statistics by adaptive input state control}
\label{sec:results}
\subsection{Suppression of probabilities by local polarization rotations}

As explained in Sec. \ref{sec:theory}, the probabilities $P(0,a)$ ($P(a,0)$) can be suppressed by destructive quantum interferences between the components $\ket{00}$ and $\ket{01}$ ($\ket{10}$). Making use of the initial quantum coherence provided by the entanglement source, the suppression can thus be optimized by searching for the polarization rotation angles $\phi_M$ with equal probabilities for the outcomes $(0,0)$ and $(0,1)$. Due to the symmetry of the state generated by the entanglement source, we can assume that $P(0,1)\approx P(1,0)$, so that the same rotation angle $\phi_M$ can be used for both photons. A single rotation angle $\phi_M$ simultaneously minimizes both $P(0,a)$ and $P(a,0)$ as can be verified after the determination of $\phi_M$ by directly measuring the residual probabilities of these outcomes. 

\begin{figure}[htbp]
  \centering
  \includegraphics[width=120mm]{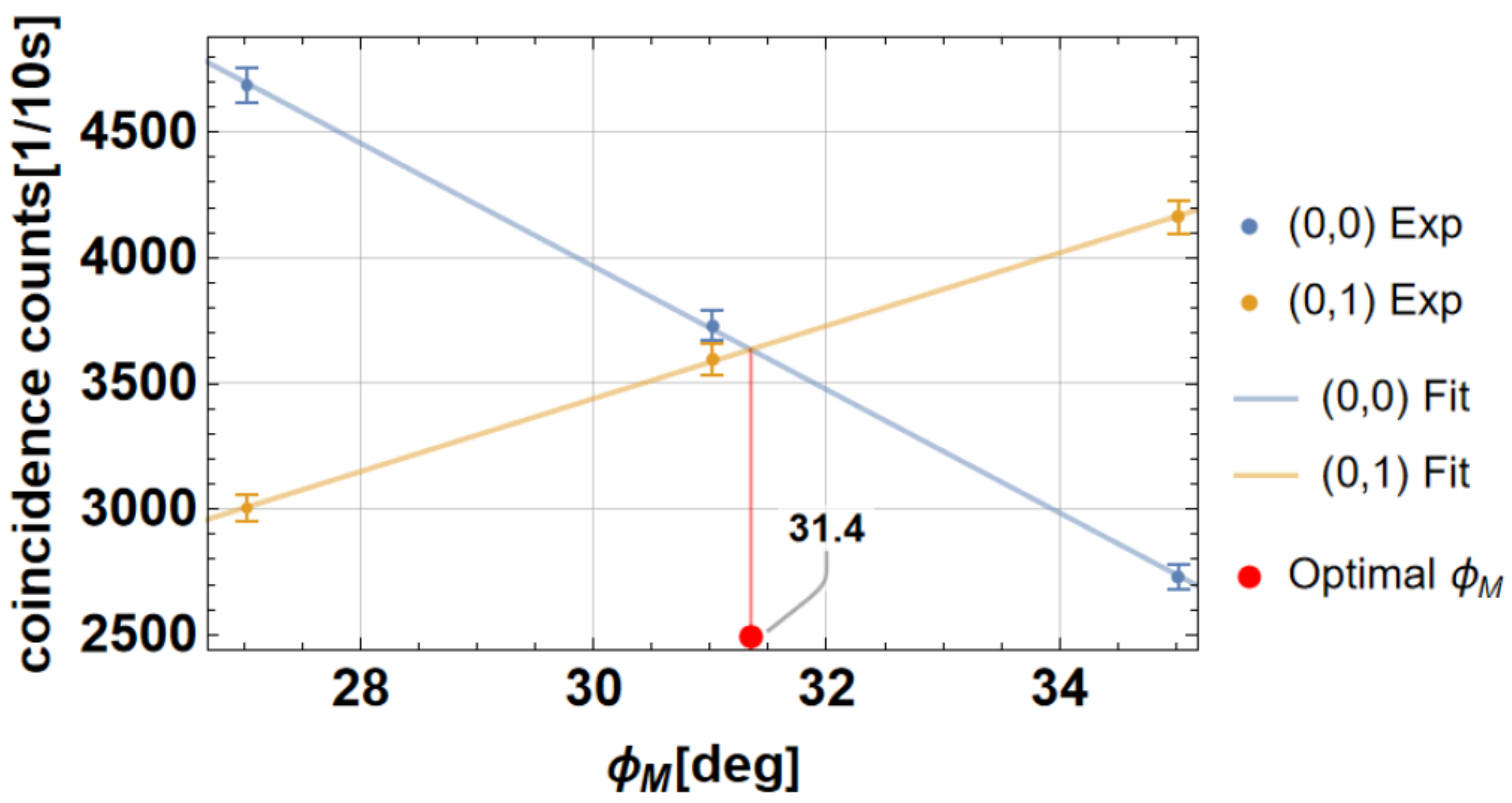}
  \caption{Determination of the optimal value of $\phi_M$ for $\phi_S = 22.5^\circ$. Count rates are obtained for the outcomes $(0,0)$ and $(0,1)$ at three different settings of $\phi_M$. The optimal value of $\phi_M$ is found at the intersection of the lines representing the linear dependence of the count rates on $\phi_M$. For $\phi_S = 22.5^\circ$, the optimal value is found at $\phi_M=31.4^\circ$. Each take three data points and are related by a linear fit; the intersection of the two lines can be the optimal $\phi_M$.}
  \label{fig:optimal phi_M}
\end{figure}

To determine the optimal value of $\phi_M$ for a specific value of $\phi_S$, we measured the coincidence counts of the outcome $(0,0)$ or $(0,1)$ at three different values of $\phi_M$ close to the theoretically expected value for that setting of $\phi_S$. Since the dependence of the coincidence counts on $\phi_M$ is approximately linear in that region, three settings of $\phi_M$ are sufficient to identify the optimal value of $\phi_M$, where the probabilities of $P(0,0)$ and $P(0,1)$ are approximately equal. Fig. \ref{fig:optimal phi_M} illustrates this method of finding the optimal values of $\phi_M$ for the case of $\phi_S = 22.5^\circ$. The procedure is relatively simple and can be used to quickly identify the optimal local polarization rotation $\phi_M$ for several different degrees of entanglement. Tab. \ref{table:obtaid phi_M} shows the results we obtained for nine different values between $\phi_M=0^\circ$ and $\phi_M=45^\circ$. 

\begin{table}[htbp]
  \centering
  \caption{Experimental results of the optimal polarization rotations $\phi_M$ for different settings of the parameter $\phi_S$. }
  \label{table:obtaid phi_M}
  \footnotesize
  \begin{tabular}{@{}cc} \hline
    $\phi_S [deg]$ & $\phi_M [deg]$ \\ \hline
    $0$ & $46.3$  \\
    $10$ & $38.8$  \\
    $17.5$ & $33.7$  \\
    $20$ & $32.7$  \\
    $22.5$ & $31.4$  \\
    $25$ & $30.5$  \\
    $27.5$ & $29.5$  \\
    $35$ & $26.7$  \\
    $45$ & $22.2$  \\ \hline
  \end{tabular} \\
\end{table}

The success of the suppression of probabilities for $P(a,0)$ and $P(0,a)$ can be confirmed by measuring the corresponding detection probabilities at the optimal settings of $\phi_M$. The results are shown in Tab. \ref{table:low limit of obtained probabilities}. The statistical average of $P(0,a)$ and $P(a,0)$ for the nine values of $\phi_S$ is $0.010 \pm 0.002$ and $0.007 \pm 0.002$ respectively, where the errors give the standard deviations for the distribution of values in the nine data points. Since the errors do not depend much on $\phi_S$, we assume that they mostly originate from the errors in the correlation of the $HV$-polarizations, $C_{HV}$. For a completely random state, $C_{HV}=0$ and $P(a,0)=P(0,a)=1/4$. Since the relations between probabilities and density matrix elements are linear, we conclude that the error in $P(a,0)$ and $P(0,a)$ caused by values of $C_{HV}$ smaller than one is given by $(1-C_{HV})/4$. These values are shown in Tab. \ref{table:low limit of obtained probabilities} alongside the values of $P(a,0)$ and $P(0,a)$. The similarity in the magnitude of errors suggests that the residual probabilities of $P(a,0)$ and $P(0,a)$ are an unavoidable consequence of the experimental imperfections that are also responsible for the errors in the correlations between the $HV$-polarizations of the emitted photon pairs.

\begin{table}[htbp]
  \centering
  \caption{Values of $P(0,a), P(a,0)$ and $\frac{1}{4}(1-C_{HV})$, which provides the lower limit of all obtainable probabilities.}
  \label{table:low limit of obtained probabilities}
  \footnotesize
  \begin{tabular}{@{}cccc} \hline
   $\phi_S [deg]$ & $P(0,a)$ & $P(a,0)$ & $\frac{1}{4}(1-C_{HV})$ \\ \hline
   0 & $0.0072\pm 0.0008$ & $0.0030\pm 0.0005$ & $0.0063\pm 0.0033$ \\
   10 & $0.0078\pm 0.0008$ & $0.0088\pm 0.0009$ & $0.0065\pm 0.0033$ \\
   17.5 & $0.0110\pm 0.0010$ & $0.0070\pm 0.0008$ & $0.0060\pm 0.0033$ \\
   20 & $0.0100\pm 0.0009$ & $0.0071\pm 0.0008$ & $0.0067\pm 0.0033$ \\
   22.5 & $0.0104\pm 0.0010$ & $0.0062\pm 0.0007$ & $0.0071\pm 0.0033$ \\
   25 & $0.0125\pm 0.0010$ & $0.0086\pm 0.0009$ & $0.0068\pm 0.0033$ \\
   27.5 & $0.0111\pm 0.0010$ & $0.0080\pm 0.0008$ & $0.0062\pm 0.0032$ \\
   35 & $0.0137\pm 0.0011$ & $0.0088\pm 0.0009$ & $0.0071\pm 0.0032$ \\
   45 & $0.0104\pm 0.0009$ & $0.0097\pm 0.0009$ & $0.0081\pm 0.0032$ \\ \hline
  \end{tabular} \\
\end{table}

\subsection{Evaluation of quantum contextuality}

\begin{figure}[htbp]
  \centering
  \includegraphics[width=100mm]{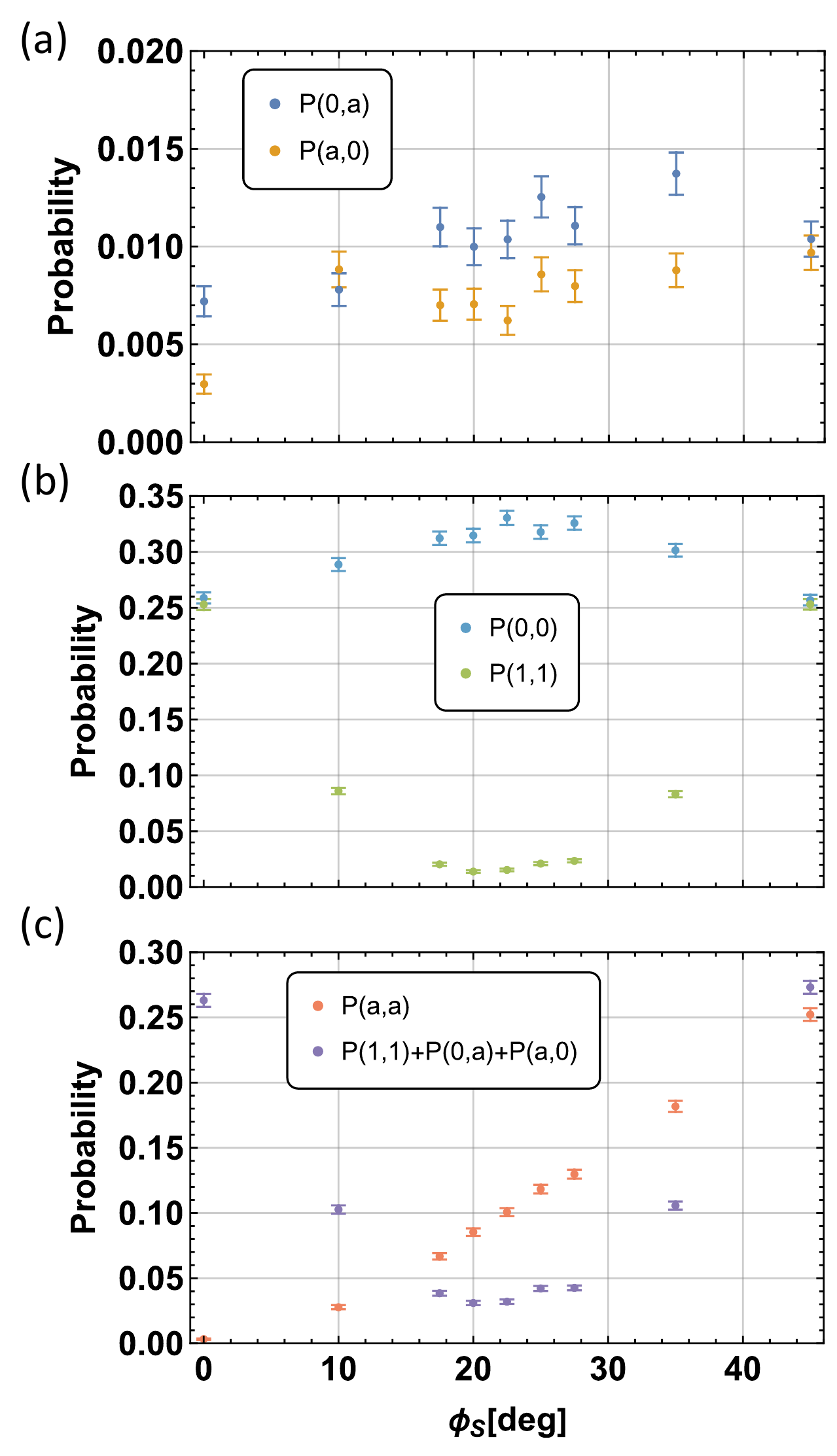}
  \caption{Experimental results for the outcome probabilities $P(0,a), P(a,0), P(1,1)$ and $P(a,a)$ as a function of the parameter $\phi_S$ that determines the degree of entanglement. $\phi_S = 0^\circ$ is the separable case and $\phi_S = 45^\circ$ is the maximally entangled state. (a) shows the results for the suppression of $P(0,a)$ and $P(a,0)$, (b) shows the results for the suppression of $P(1,1)$ in comparison with $P(0,0)$, and (c) shows the results for  $P(a,a)$ and $P(0,a) + P(a,0) + P(1,1)$. Quantum contextuality is observed when $P(a,a)$ is larger than the sum of the three suppressed probabilities. See the supplemental material for the raw data obtained in the experiment.}
  \label{fig:four probabilities involved in the criterion}
\end{figure}

The purpose of adaptive state control is the observation of quantum contextuality. By suppressing the probabilities of the outcomes $(0,a)$, $(a,0)$ and $(1,1)$, we should obtain a much higher probability of the outcome $(a,a)$, resulting in a clear violation of the inequality given in Eq.(\ref{eqn:criterion of non-contexual logic}). This violation of non-contextual logic can be observed directly in the count rates of each of the outcomes. The raw data we obtained for the different settings is shown in Tab. \ref{table:raw data of the four measurement outcomes}. The number of counts $N(a,a)$ is higher than the sum of the other three counts for all settings from $\phi_S = 17.5^\circ$ to $ \phi_S = 35^\circ$. The variation of the degree of entanglement results in low counts in the outcome $(1,1)$ with a minimum between $\phi_S = 20^\circ$ and $ \phi_S = 22.5^\circ$. Fig. \ref{fig:four probabilities involved in the criterion} shows the corresponding probabilities. Fig. \ref{fig:four probabilities involved in the criterion} (a) shows the suppression of $P(0,a)$ and $P(a,0)$ discussed previously. Values are close to $0.01$ or one percent throughout, confirming the successful optimization of $\phi_M$. The symmetry breaking of the quantum state creates a difference between $P(0,a)$ and $P(a,0)$. Fig. \ref{fig:four probabilities involved in the criterion} (b) shows the suppression of $P(1,1)$ by intermediate values of $\phi_S$. Note the parabolic profile of $P(1,1)$ around the minimum between $\phi_S = 20^\circ$ and $\phi_S = 22.5^\circ$. The probability $P(0,0)$ is shown for comparison. Since the optimization of $\phi_M$ required that $P(0,0)$, $P(0,1)$, and $P(1,0)$ are approximately equal, the increase of $P(0,0)$ is equal to about one third of the decrease in $P(1,1)$. Finally, Fig. \ref{fig:four probabilities involved in the criterion} (c) illustrates the observation of quantum contextuality by comparing the probability $P(a,a)$ with the sum of the suppressed probabilities, $P(0,a) + P(a,0) + P(1,1)$. Due to the dependence of the suppression of $P(1,1)$ on $\phi_S$, the sum of the suppressed probabilities has a minimum between $\phi_S = 20^\circ$ and $\phi_S = 22.5^\circ$. On the other hand, $P(a,a)$ increases monotonically with $\phi_S$. As a result, quantum contextuality can be observed clearly at higher values of $\phi_S$. Contextuality only disappears close to the maximally entangled state at $\phi_S = 45^\circ$.

\begin{table}[htbp]
  \centering
  \caption{Counts for the four outcomes $(0,a), (a,0), (1,1)$ and $(a,a)$ per ten seconds.}
  \label{table:raw data of the four measurement outcomes}
  \footnotesize
  \begin{tabular}{@{}ccccc} \hline
   $\phi_S [deg.]$ & $N(0,a)$ & $N(a,0)$ & $N(1,1)$ & $N(a,a)$ \\ \hline
   0 & 88 & 36 & 3354 & 40 \\
   10 & 88 & 95 & 954 & 304 \\
   17.5 & 125 & 79 & 231 & 752 \\
   20 & 113 & 79 & 155 & 959 \\
   22.5 & 118 & 71 & 172 & 1148 \\
   25 & 145 & 98 & 240 & 1357 \\
   27.5 & 136 & 97 & 285 & 1586 \\
   35 & 163 & 106 & 1002 & 2153 \\
   45 & 136 & 123 & 3513 & 3430 \\ \hline
  \end{tabular} \\
  \end{table}

It may be interesting to consider the role of entanglement in the simultaneous suppression of the probabilities $P(0,a)$, $P(a,0)$, and $P(1,1)$, and in the subsequent observation of a non-zero value of $P(a,a)$. Fig. \ref{fig:four probabilities involved in the criterion} (c) clearly shows that $P(a,a)$ increases together with the amount of entanglement characterized by the witness $W_E$. Since the maximal value of $P(a,a)$ is $0.25$, this increase represents an increase in the randomness of the relation between the outcomes of $\{a,b\}$ measurements for the two photons. Local polarizations always prefer outcomes of $b$ and therefore limit the probability $P(a,a)$. On the other hand, maximal entanglement makes it impossible to suppress $P(1,1)$ together with $P(0,a)$ and $P(a,0)$, since correlations between the outcomes of $\{0,1\}$ measurements necessarily weaken the correlations between $\{0,1\}$ and $\{a,b\}$. Optimal quantum contextuality is obtained when the suppression of $P(0,a)$ and $P(a,0)$ is achieved by a combination of local polarization suppressing all $a$-results and entanglement correlations suppressing the outcome combinations $(0,a)$ and $(1,b)$. This combination can apply simultaneously to a local suppression of $1$-results and the combinations $(1,1)$ and $(0,0)$. Simultaneous suppression of all three outcomes $P(0,a)$, $P(a,0)$, and $P(1,1)$ can be achieved by both local polarization and entanglement, but only a balanced combination of both can achieve an optimal suppression effect. At the same time, suppression of $a$-results by the local polarization will also limit the probability $P(a,a)$, adding an additional cost to the use of local polarization in the suppression of probabilities.

\section{Contrast between probabilities}
\label{sec:discussion}

\begin{figure}[htbp]
  \centering
  \includegraphics[width=120mm]{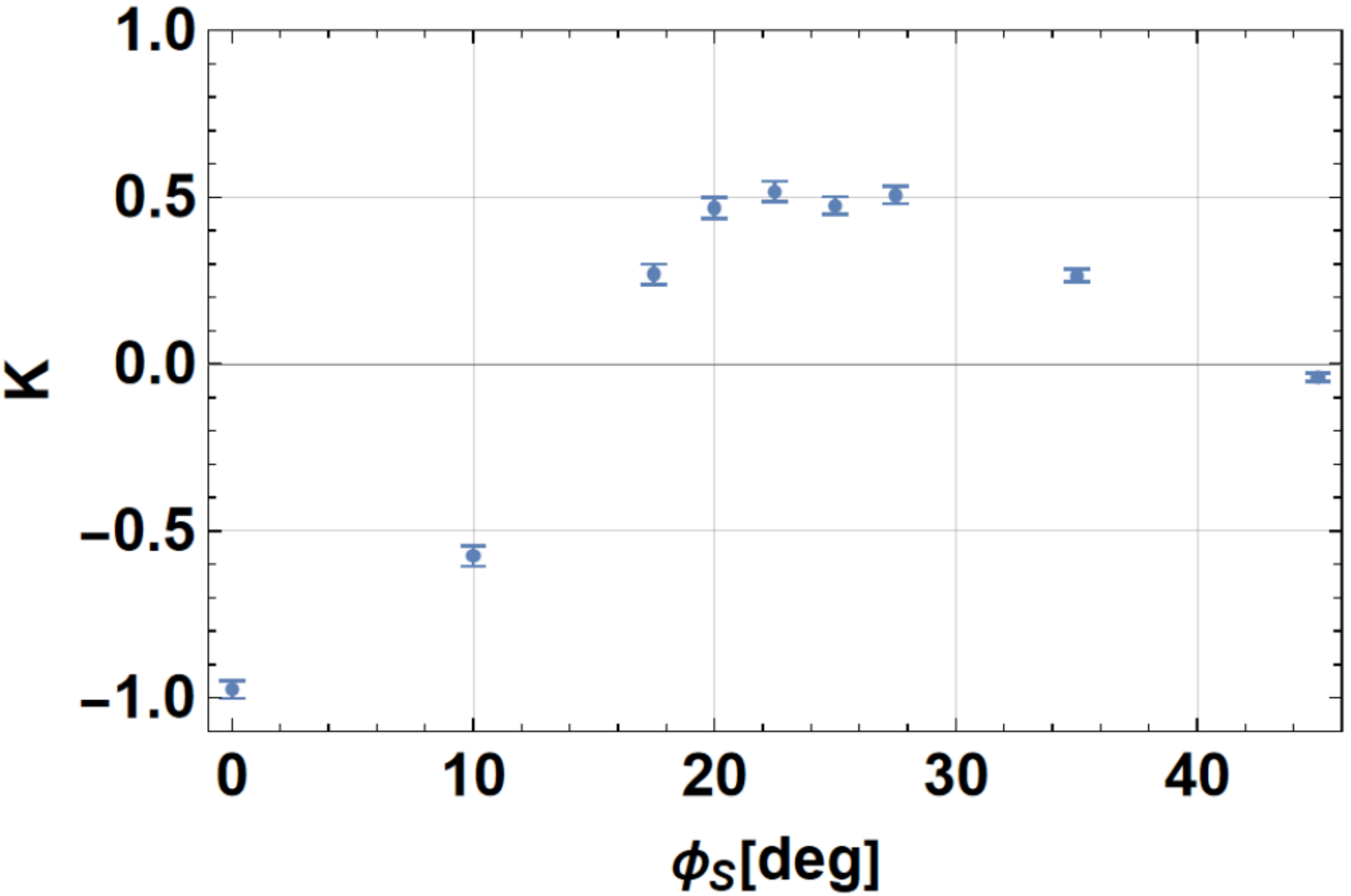}
  \caption{Experiment result of the contrast function defined by Eq. (\ref{eqn:def of contrast function}) against $\phi_S$.}
  \label{fig:experimental result of K}
\end{figure}

The probability $P(1,1)$ can only be suppressed by varying the entanglement of the input state. The lowest value of $P(1,1)$ determined from the experimental data was obtained at $\phi_S = 20^\circ$ with $P(1,1)=0.0140 \pm 0.0011$. The data shows a relatively flat minimum with very similar values of $P(1,1)$ between $\phi_S = 20^\circ$ and $\phi_S = 22.5^\circ$. Moreover, the error rate is consistent with the two-photon visibilities $C_{HV}$ and $C_{PM}$ of the source. The sum of the suppressed probabilities at $\phi_S = 20^\circ$ is $P(0,a)+P(a,0)+P(1,1) = 0.0311 \pm 0.0016$. Experimental errors thus make it impossible to observe a probability of less than three percent in the suppressed probabilities. To observe quantum contextuality, it is therefore necessary to obtain a sufficiently high value of $P(a,a)$. At $\phi_S = 20^\circ$, this value is at $P(a,a)=0.0854 \pm 0.0028$. However, higher values are obtained as $\phi_S$ increases. It is therefore desireable to evaluate the contrast between the low value of the probability sum $P(0,a)+P(a,0)+P(1,1)$ and the high value of the probability $P(a,a)$. This contrast can be defined as
\begin{eqnarray}
  K \equiv \frac{P(a,a) - P(0,a) - P(a,0) - P(1,1)}{P(a,a) + P(0,a) + P(a,0) + P(1,1)}.
  \label{eqn:def of contrast function}
\end{eqnarray}
By definition, the contrast $K$ is one if and only if $P(0,a), P(a,0)$ and $P(1,1)$ are all zero. If the probabilities do not violate the inequality in Eq. (\ref{eqn:criterion of non-contexual logic}), $K$ has a negative value. The value of $K$ at $\phi_S = 20^\circ$ is $K=0.467\pm 0.032$, a clear violation of the inequality and a sufficiently clear contrast between the suppressed probabilities and the probability $P(a,a)$. However, the increase in $P(a,a)$ now results in slightly larger values of $K$ for $\phi_S = 22.5^\circ$, $\phi_S = 25^\circ$, and even $\phi_S = 27.5^\circ$. The different values of $K$ are shown in Fig. \ref{fig:experimental result of K} and in Tab. \ref{table:result of contrast function}. The result clearly indicate that the observation of quantum contextuality is easier when the entanglement is larger than the entanglement of the ideal state defined by Eqs. (\ref{eqn:condition one})-(\ref{eqn:condition three}).

  \begin{table}[htbp]
  \centering
  \caption{Contrast $K$ achieved at different settings of $\phi_S$.}
  \label{table:result of contrast function}
  \footnotesize
  \begin{tabular}{@{}cc} \hline
   $\phi_S [deg]$ & $K$ \\ \hline
   0 & $-0.975\pm 0.026$ \\
   10 & $-0.575\pm 0.031$ \\
   17.5 & $0.269\pm 0.031$  \\
   20 & $0.467\pm 0.032$  \\
   22.5 & $0.518\pm 0.030$ \\
   25 & $0.475\pm 0.027$  \\
   27.5 & $0.506\pm 0.026$  \\
   35 & $0.265\pm 0.019$  \\
   45 & $-0.040\pm 0.013$ \\ \hline
  \end{tabular} \\
  \end{table}

As discussed before, the maximal value of the contrast $K$ is obtained as the result of a trade-off between the contributions of local polarization and entanglement to the suppression of $P(1,1)$ and the increase of $P(a,a)$. The condition imposed by the simultaneous suppression of $P(a,0)$ and $P(0,a)$ limits the contribution of entanglement to the suppression of $P(1,1)$, requiring a local suppression of the outcomes $a$. However, this local suppression of $a$ also limits the probability $P(a,a)$. Adaptive input state control optimizes these relations in the presence of experimental imperfections, making optimal use of the available local polarization and the available entanglement. Although the contrasts that can be achieved are limited by the experimental imperfections, the range of parameters for which contrasts close to the maximum value can be obtained is much larger than the ideal case would seem to suggest. Adaptive input state control thus optimizes the robustness of quantum contextuality against experimental errors and decoherence. 

\section{Conclusions}
\label{sec:conclusion}

Quantum state preparation can be optimized to achieve maximal quantum contextuality in the presence of experimental imperfections. We achieve this by defining state preparation in terms of output probabilities that need to be minimized in order to observe quantum contextuality. In the present scenario, it was necessary to first determine the degree of entanglement of a two-photon state by setting a control parameter $\phi_S$, followed by an optimization of local polarization rotations $\phi_M$. Significantly, we were able to achieve a highly efficient optimization of $\phi_M$ by making use of the fact that quantum interference is maximal when the amplitudes of the interfering components are equal. Using this optimization method, we were then able to optimize the trade-off between entanglement and local polarization described by $\phi_S$. The results show that the same contrast $K$ between the probability of the paradoxical outcome $(a,a)$ and the suppressed probabilities can be obtained for a wide range of different degrees of entanglement. This indicates that entanglement can partially compensate the effects of errors by increasing the probability of the outcomes $(a,a)$. Importantly, adaptive state control allows us to directly optimize the statistics that characterize the actual input state, without any theoretical assumptions about the quantum coherence of the source. 

As mentioned in the introduction, adaptive input state control has three distinct advantages. (a) It allows us to partially compensate the effects of experimental imperfections by optimizing the parameters that control the state preparation process, improving the performance of noisy sources. (b) It allows us to directly relate quantum state preparation to specific correlations relevant for the observation of contextuality paradoxes or similar purposes. (c) The optimization of quantum interference effects can be realized by observing the output probabilities of the components that interfere with each other, providing a particularly simple method of optimizing quantum coherent effects.

In conclusion, adaptive input state control allows us to optimize the non-classical properties of quantum states in the presence of experimental imperfections by directly implementing specific statistical correlations that are relevant for the intended effects. Adaptive input state control thus makes fundamental aspects of quantum systems more accessible and may facilitate their application in future quantum information technologies.

\begin{backmatter}

\bmsection{Acknowledgments}
This work was supported by JST SPRING, Grant Number JPMJSP2132.

\bmsection{Disclosures}
The authors declare no conflicts of interest.

\bmsection{Data availability} See the supplemental material.

\end{backmatter}

\appendix
\section{Supplemental material: Complete set of raw data obtained in the experiment}
In this supplementary material, we include the raw data obtained in the experiment: Tab. \ref{tab:raw1} shows the counts for 10 seconds used in the results presented in Fig. 2; Tab. \ref{tab:raw2} shows the raw data used to derive $P(0,a), P(a,0), P(1,1)$ and $P(a,a)$; The numbers in Tab. \ref{tab:raw2} are also 10-second counts.

\renewcommand{\thetable}{A\arabic{table}}
\setcounter{table}{0}
\begin{table}[htbp]
\centering
\caption{Absolute counts in $\lbrace H,V \rbrace$ or $\lbrace P,M \rbrace$ basis for each $\phi_S$. All counts were obtained from a 10-second measurement.}
\begin{tabular}{ccccccccc}
\hline
   $\phi_S[deg]$ & $N(H,H)$ & $N(H,V)$ & $N(V,H)$ & $N(V,V)$ & $N(P,P)$ & $N(P,M)$ & $N(M,P)$ & $N(M,M)$ \\
   \hline
   0 & 11326 & 119 & 27 & 47 & 3013 & 2926 & 2836 & 3033 \\
   10 & 10379 & 115 & 26 & 379 & 1895 & 3657 & 3611 & 1853 \\
   17.5 & 10030 & 107 & 28 & 1101 & 1369 & 4353 & 4387 & 1342 \\
   20 & 9860 & 111 & 40 & 1336 & 1152 & 4711 & 4506 & 1182 \\
   22.5 & 9391 & 109 & 51 & 1666 & 1082 & 4655 & 4640 & 969 \\
   25 & 9218 & 108 & 45 & 1947 & 946 & 4854 & 4935 & 895 \\
   27.5 & 9695 & 105 & 46 & 2415 & 767 & 5327 & 5464 & 812 \\
   35 & 7923 & 114 & 54 & 3689 & 459 & 5651 & 5611 & 430 \\
   45 & 5902 & 109 & 81 & 5611 & 207 & 5629 & 5795 & 175 \\
   \hline
\end{tabular}
  \label{tab:raw1}
\end{table}

\begin{table}[htbp]
\centering
\caption{Absolute counts with respect to $(F,F)$, $(F,W)$, $(W,F)$ and $(W,W)$ for each $\phi_S$. All counts were obtained from a 10-second measurement.}
\begin{tabular}{ccccccccc}
\hline
   $\phi_S[deg]$ & $N(0,0)$ & $N(0,1)$ & $N(1,0)$ & $N(1,1)$ & $N(0,a)$ & $N(0,b)$ & $N(1,a)$ & $N(1,b)$ \\
   \hline
   0 & 3431 & 2989 & 3484 & 3354 & 88 & 5933 & 81 & 6119 \\
   10 & 3201 & 3377 & 3556 & 954 & 88 & 6510 & 564 & 4120 \\
   17.5 & 3527 & 3560 & 3979 & 231 & 125 & 7071 & 1442 & 2727 \\
   20 & 3496 & 3465 & 3993 & 155 & 113 & 7114 & 1782 & 2299 \\
   22.5 & 3688 & 3638 & 3661 & 172 & 118 & 7240 & 2159 & 1862 \\
   25 & 3624 & 3573 & 3967 & 240 & 145 & 7293 & 2492 & 1635 \\
   27.5 & 3952 & 3917 & 3977 & 285 & 136 & 7723 & 2990 & 1440 \\
   35 & 3631 & 3609 & 3803 & 1002 & 163 & 7020 & 4105 & 584 \\
   45 & 3565 & 3394 & 3408 & 3513 & 136 & 6444 & 6401 & 114 \\
   \hline
\end{tabular}
  \label{tab:raw2}

  \begin{tabular}{ccccccccc}
\hline
   $\phi_S[deg]$ & $N(a,0)$ & $N(a,1)$ & $N(b,0)$ & $N(b,1)$ & $N(a,a)$ & $N(a,b)$ & $N(b,a)$ & $N(b,b)$ \\
   \hline
   0 & 36 & 38 & 6223 & 5824 & 40 & 38 & 95 & 12072 \\
   10 & 95 & 417 & 6453 & 3788 & 304 & 144 & 292 & 10224 \\
   17.5 & 79 & 1353 & 7323 & 2519 & 752 & 608 & 774 & 9123 \\
   20 & 79 & 1616 & 7264 & 2237 & 959 & 756 & 870 & 8649 \\
   22.5 & 71 & 2086 & 7501 & 1748 & 1148 & 959 & 1098 & 8196 \\
   25 & 98 & 2380 & 7434 & 1511 & 1357 & 1112 & 1210 & 7791 \\
   27.5 & 97 & 2850 & 7842 & 1359 & 1586 & 1357 & 1579 & 7697 \\
   35 & 106 & 4083 & 7199 & 671 & 2153 & 1964 & 2037 & 5691 \\
   45 & 123 & 6317 & 6107 & 144 & 3430 & 3370 & 3318 & 3483 \\
   \hline
\end{tabular}
  \label{tab:raw3}
  
\end{table}

\newpage


\bibliography{sample}

\end{document}